\documentclass[lettersize,journal]{IEEEtran}
\usepackage{amsmath,amsfonts}
\usepackage{algorithmic}
\usepackage{algorithm}
\usepackage{array}
\usepackage[caption=false,font=normalsize,labelfont=sf,textfont=sf]{subfig}
\usepackage{textcomp}
\usepackage{stfloats}
\usepackage{url}
\usepackage{verbatim}
\usepackage{graphicx}
\usepackage{cite}
\usepackage{xspace}
\usepackage{inconsolata}
\usepackage{multirow}
\usepackage{colortbl}
\usepackage{amsfonts}
\usepackage{siunitx}
\usepackage{makecell}
\usepackage{xcolor}
\usepackage{xurl}
\usepackage{hyperref}
\usepackage{enumitem}
\usepackage{wrapfig}
\usepackage{booktabs}
\usepackage{pifont}

\newcommand{\jn}[1]{{\color{black}{#1}}}

\newcommand{\xmark}{\textcolor{red}{\ding{55}}}

\newcommand{\q}[1]{{\textit{``#1"}}}

\newcommand{\eg}{\emph{e.g.}\xspace}
\newcommand{\ie}{\emph{i.e.}\xspace}
\newcommand{\stab}{\vspace{1.2ex}\noindent}
\newcommand{\stitle}[1]{\stab\noindent{\bf #1}}

\newcommand*{\img}[1]{%
      \raisebox{-.1\height}{%
        \includegraphics[
          height=\baselineskip,
          keepaspectratio,
        ]{#1}%
      }%
}
\NewDocumentCommand{\yuyu}{ mO{} }{\textcolor{red}{\textsuperscript{\textit{Yuyu}}\textsf{\textbf{\small[#1]}}}}

\newcommand{\tool}{\emph{VisTR}\xspace}

\usepackage[normalem]{ulem}
\definecolor{mred}{rgb}{.80,.12,.30}
\definecolor{grey}{rgb}{0.5,0.5,0.5}
\definecolor{lgrey}{rgb}{0.7,0.7,0.7}
\definecolor{purple}{rgb}{.75,0,.85}
\definecolor{pistachio}{rgb}{0.58, 0.77, 0.45}
\definecolor{myorange}{rgb}{0.94, 0.36, 0.13}

\begin{document}

\title{\tool: Visualizations as Representations for \\ Time-series Table Reasoning}

\author{Jianing Hao, Zhuowen Liang, Chunting Li, Yuyu Luo, Jie Li, Wei Zeng, \IEEEmembership{Member,~IEEE}
\thanks{J. Hao, C. Li, Z. Liang, Y. Luo, and W. Zeng are with the Hong Kong University of Science and Technology (Guangzhou). Y. Luo and W. Zeng are also with the Hong Kong University of Science and Technology Email: \{jhao768@connect., cli087@connect., zliang321@connect., yuyuluo@, weizeng@\}hkust-gz.edu.cn}%
\thanks{J. Li is with the Tianjin University. jie.li@tju.edu.cn}%
\thanks{W. Zeng is the corresponding author.}}

\markboth{Journal of \LaTeX\ Class Files,~Vol.~14, No.~8, August~2021}%
{Hao \MakeLowercase{\textit{et al.}}: \tool: Visualizations as Representations for Time-series Table Reasoning}

\IEEEpubidadjcol

\maketitle

\begin{abstract}
Time-series table reasoning interprets temporal patterns and relationships in data to answer user queries.
Despite recent advancements leveraging large language models (LLMs), existing methods often struggle with pattern recognition, \jn{context loss} in long time-series data, and the lack of visual-based reasoning capabilities.
To address these challenges, we propose \tool, a framework that places visualizations at the core of the reasoning process.
By transforming tables into fixed-size visualization references, \tool captures key trends, anomalies, and temporal relationships, facilitating intuitive and interpretable reasoning. 
These visualizations are aligned with user input, \ie, charts, text, and sketches, through a fine-tuned multimodal LLM, ensuring robust cross-modal alignment. 
To handle large-scale data, \tool integrates pruning and indexing mechanisms for scalable and efficient retrieval. 
Finally, an interactive visualization interface supports seamless multimodal exploration, enabling users to interact with data through both textual and visual modalities.
Quantitative comparisons with time-series matching methods and multimodal LLMs demonstrate the effectiveness of \tool in aligning multimodal inputs, improving reasoning accuracy, and reducing latency.
Furthermore, representative usage scenarios and user feedback validate \tool's applicability to various time-series reasoning and exploration tasks.
\end{abstract}


\section{Introduction}

\begin{figure*}[t]
	\centering
	\includegraphics[width=0.99\linewidth]{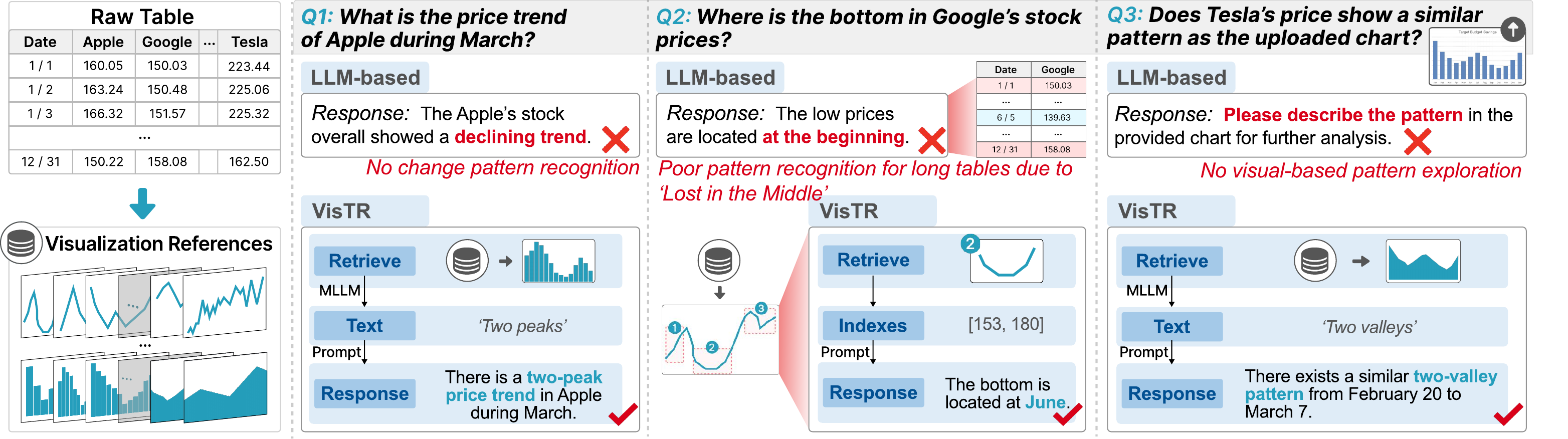}
	\vspace{-1mm}
	\caption{ 
		\tool leverages visualizations as the {\em representations} to bridge the gap between time-series data and user queries.
		In this way, \tool enhances existing LLM-based table reasoning methods by enabling robust data change pattern recognition (Q1), improving pattern recognition for long-term series (Q2), and facilitating visual-based pattern exploration (Q3). 
	}
	\vspace{-1mm}
	\label{fig:teaser}
\end{figure*}

\IEEEPARstart{T}{ime}-series tabular data are ubiquitous, forming the backbone of decision-making processes across domains such as finance, healthcare, and logistics\cite{DBLP:journals/tkde/ChaiWLNL23}.
These datasets record sequential information over time, enabling the discovery of temporal trends and relationships that drive actionable insights\cite{pattern_rec_2012, spade_2006}. 
Time-series table reasoning, a key task in this context, focuses on interpreting these patterns to answer user queries, such as \q{What is the price trend of Apple during March?} (see Figure~\ref{fig:teaser}-Q1).
Effectively addressing this task requires a comprehensive understanding of not only the raw tabular data but also the temporal trends, interrelationships among data points, and alignment with the context of user queries.

Substantial research efforts have been dedicated to automated table reasoning, aiming to generate the corresponding answer to the question following the user requirement according to the provided table\cite{TR_survey_2025}.
Unlike general natural language reasoning\cite{NLR_2024}, table reasoning requires both semantic understanding and structural comprehension of tables.
Recently, large language models (LLMs) have demonstrated promising capabilities in table reasoning tasks\cite{DBLP:journals/pvldb/LiLCLT24, DBLP:journals/corr/abs-2408-05109, DBLP:journals/corr/abs-2410-10762}. 
These solutions rely heavily on text-based reasoning, which however struggles to emulate human-like cognitive processes.

This limitation often leads to failures in tasks requiring complex pattern recognition and the alignment with user intent.
Some core limitations are exemplified in Figure~\ref{fig:teaser}.
1)~\textit{Limited Pattern Recognition.}
LLMs often struggle to identify and interpret temporal data patterns, such as trends, peaks, or valleys\cite{Finflier_2024, sql-nl-survey}.
For example, in Figure~\ref{fig:teaser}-Q1, an LLM misinterprets Apple’s stock prices as exhibiting a \q{declining trend} rather than recognizing the correct two-peak trend.
2)~\textit{Challenges with Long Time-series Data.} LLMs often lose focus and context when reasoning over long time-series data, leading to errors caused by \jn{\q{lost-in-the-middle} issues}\cite{lostmiddle_2024}. 
In Figure~\ref{fig:teaser}-Q2, an LLM erroneously identifies Google’s stock price bottom as occurring \q{at the beginning} rather than pinpointing the correct date.
3)~\textit{Inability to Perform Visual-based Reasoning.}
Existing LLMs cannot effectively reason over external visual references, limiting their ability to align time-series data with visual patterns\cite{DBLP:journals/corr/abs-2406-07815}.
For example, in Figure~\ref{fig:teaser}-Q3, an LLM fails to determine whether Tesla’s stock price exhibits a similar pattern to a user-provided chart, demonstrating the gap in visual-based reasoning.


To address these limitations, recent studies have sought to leverage visualizations (e.g., line and bar charts) to enhance downstream tasks in time-series table reasoning, such as long-term forecasting~\cite{visionts_2024,timevlm_2025} and classification~\cite{image_class_2025}.
This approach is beneficial because visual representations enable models to utilize image-based techniques for uncovering intricate temporal patterns, including trends, outliers, seasonality, and relationships~\cite{chatts_2024,plotsts_2024}. 
Beyond improving model performance, visualizations also serve as an intuitive cognitive bridge, allowing analysts to discover patterns that may remain hidden in raw tabular data.

In line with this research direction, we introduce a new framework that integrates visualizations as \emph{representations} into multimodal LLMs (MLLMs).
The \textbf{key idea} of our framework is to transform {\em time-series data} into a set of meaningful and insightful visualizations, termed {\em visualization references}. These references act as concise representations of the data, capturing essential insights such as trends, anomalies, and temporal relationships, while aligning closely with user intentions for reasoning tasks.
To realize this idea, our framework consists of four integrated modules: \emph{visualization referencing} for transforming tables into meaningful visualization references; \emph{visualization pruning} for filtering less informative visualizations; \emph{visualization alignment} for aligning visualization references with user intentions through a fine-tuned MLLM; \emph{visualization interaction} for enabling intuitive multimodal exploration.

Our framework offers several key advantages over existing LLM-based table reasoning methods.

\textit{(1) Enhanced Pattern Recognition Across Time Scales.}
Visualization references help MLLMs quickly identify both short- and long-term data change patterns that might be missed in raw tabular data.
For example, in Figure~\ref{fig:teaser}-Q1, the \q{two-peak} price trend of Apple during March is clearly represented through the bar chart retrieved from \tool's database.

\textit{(2) Mitigating \jn{Context Loss} in Long Time-series.} 
Fixed-size visualization references condense data patterns, mitigating context loss issues in long time-series reasoning. 
These \jn{fixed-size visualization references} are generated with a consistent aspect ratio to prevent shape distortion.
For example, in Figure~\ref{fig:teaser}-Q2, a year-long stock price change is condensed into a single visualization, making it easy to pinpoint the price bottom without losing contextual accuracy.

\textit{(3) Intuitive Multimodal Interaction.}
Integrating visualizations as representations enables analysts to explore the data naturally by aligning with human cognitive processes. 
For example, Figure~\ref{fig:teaser}-Q3 shows how combining textual descriptions with charts improves query interpretation and reduces reliance on raw time-series data~\cite{Lee_2020_visualquery}.

In summary, we make the following contributions:

\begin{itemize}
	\item {\bf Framework}. We propose \tool that leverages visualizations as \emph{representations} to enhance time-series table reasoning. This new framework enables robust pattern recognition and intuitive exploration for time-series data.
	
	\item {\bf Multimodal Visualization Alignment}.  We fine-tune an effective MLLM that learns a joint embedding space across three modalities, \ie, charts, text, and hand-drawn sketches. This alignment bridges user intentions with visualization references, enabling users to explore time-series tables through multimodal interactions.

	\item {\bf Prototype Construction}.  We implement a prototype system that facilitates reasoning and exploration of large-scale time-series tables.
		
    \item {\bf Evaluation.} We conduct extensive evaluations, including quantitative experiments and usage scenarios, to demonstrate the effectiveness and usability of \tool. 
    All datasets, fine-tuned models, anonymized user study data, and evaluation results are available at \url{https://github.com/HKUST-CIVAL/VisTR}.
    
\end{itemize}
\section{Related Work}
\subsection{Table Reasoning}
Table reasoning bridges free-form \jn{natural language} (NL) questions and structured tables, enabling lay users to explore tabular data effortlessly and \jn{supporting} complex tasks such as fact verification\cite{tabfact_2020} and question answering\cite{TQA_survey_2022}.
Existing LLM-based table reasoning methods can be categorized into \emph{generic table reasoning} and \emph{program-aided table reasoning}\cite{chain_of_table}.
Generic reasoning methods employ LLMs to interpret and respond to free-form NL queries by generating answers directly from tables\cite{zero_table_2023}.
However, due to the complexity of table structure, these methods struggle with large tables and offer limited explainability because of their end-to-end nature.
Program-aided table reasoning utilizes executable query languages, such as SQL, to form a structured and accountable bridge between user queries and the derived answers\cite{table_sql_2020, TAPEX_sql_2022, omnitab_sql_2022}.
Recent research has mostly focused on leveraging LLMs to transform complex user utterances into SQL queries with data context awareness\cite{dater_2023}, using prompt engineering\cite{zero_shot_2022}, chain-of-thought\cite{chain_of_table,dater_2023}, and pre-training techniques\cite{proton_2022}.
However, program-aided approaches encounter intricate limitations due to their reliance on query language.
First, SQLs primarily cater to numerical queries from tables but lack support for analytical functionality\cite{sql-nl-survey}, overlooking data change patterns.
Second, when confronted with large tables, LLMs may encounter \q{\jn{lost-in-the-middle}} issues\cite{lostmiddle_2024} \jn{when} accessing information in the middle of long contexts.
Moreover, solely textual inputs limits the expression of non-verbal intentions, such as hand-drawn sketches, reducing the flexibility of user interactions.

To overcome these limitations, this work introduces a novel framework that leverages visualizations as \emph{representations} between tables and user intentions.
Our inspiration is that visualizations can depict intuitive data patterns and illustrate complex underlying processes\cite{few2004show}, which aids swift and effective interpretation by users\cite{vis_2015}.
However, this approach also presents challenges to be tackled.
First, tabular data typically encompasses numerous features, leading to a plethora of potential visualizations\cite{sen_2021,sun2022learning} that complicate interactive interactions.
Second, there are no direct approaches available to connect free-form user queries with visualizations.

\begin{figure*}[t]
    \centering
    \includegraphics[width=0.99\linewidth]{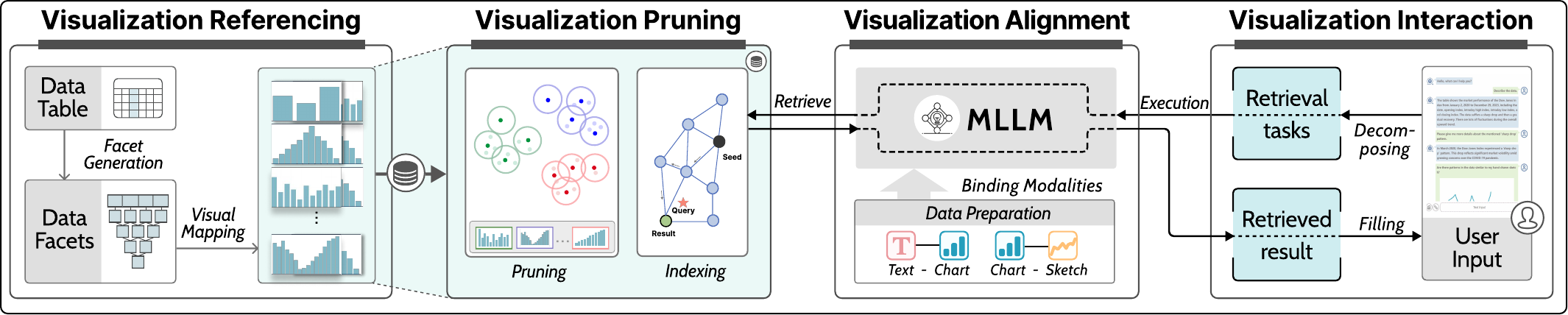}
    \vspace{-2mm}
    \caption{An overview of the proposed visualization-as-representation framework, which mainly contains four modules of
    \emph{visualization referencing}, \emph{visualization pruning}, \emph{visualization alignment}, and \emph{visualization interaction}.}
    \vspace{-3mm}
    \label{fig:pipeline}
\end{figure*}

\subsection{Multimodal Visualization Alignment}
Given that visualizations are inherently stored as images, we opt to leverage MLLMs to \jn{interpret} visualization images.
In recent years, advancements have been made in utilizing MLLMs for chart-based tasks, such as CQA~\cite{masry2022chartqa,zeng_chartqa,hoque2022chart}, chart retrieval~\cite{xiao_2022_wytiwyr}, and generation~\cite{schetinger2023doom,xiao_2023_chartspark}.
However, these models mainly support text-chart interactions, lacking the capability for cross-modal interactions involving multiple modalities such as text, chart, and sketch preferred by laymen~\cite{qetch_2018}.
Furthermore, these models are not specifically designed for effectively recognizing data change patterns in time-series data (see Sect.~\ref{ssec:evaluation}), a gap that this work seeks to fill in the context of time-series table reasoning.

Most existing multimodal data alignment methods rely on contrastive language-image pre-training (CLIP)~\cite{CLIP_2021}, which has demonstrated its adaptability and proficiency across various tasks\cite{clip_label_2022,clip_enhance_2023, clipdetect_2022}.
However, training an effective CLIP-based alignment model for visualizations is non-trivial.
First, this task involves aligning more than two modalities, including text, hand-drawn sketches, and various types of charts, whereas the original CLIP model was designed to support text-image alignment.
To encompass more modalities, we adopt joint embedding strategies that leverage charts as naturally aligned supervision and emergent alignments to unify other modalities\cite{imagebind_2023}.
Second, the effectiveness of CLIP is largely attributed to its extensive training on vast (natural) image-text datasets, whereas there is a lack of suitable training datasets for the task in this work.
Efforts have been made to construct datasets for chart-text (\eg,\cite{chart-text,masry2022chartqa}), chart-chart (\eg,\cite{scatternet_2018, peax_2020, visatlas_2022}), chart-sketch (\eg,\cite{qetch_2018,zenvisage_2017,linenet_2023}) similarity measurement.
However, these different modalities are not inherently unified in one model, and many of these datasets are not specifically designed for recognizing data change patterns.
To address this gap, we utilize data augmentation and user labeling strategies to supplement existing datasets in training the multimodal visualization alignment model.

\subsection{Visualization Storage, Indexing, and Retrieval}
To comprehensively represent tabular data, our approach will generate a large number of visualization references and encode them into vector representations, thus raising new challenges for efficiently storing, indexing, and retrieving them. 
Recent advancements in Vector Database Management Systems (VDBMS)\cite{vector_survey_2023} offer a promising solution to these challenges, as they are specifically designed for managing high-dimensional data, such as embeddings, with sophisticated storage, indexing, query processing, and query optimization capabilities\cite{DBLP:conf/acm/AsaiMZC23}.
Their ability to perform efficient similarity searches in high-dimensional spaces makes them particularly suited for tasks involving the retrieval of embeddings based on content or semantic similarity. This supports various applications such as information retrieval, e-commerce, and recommendation platforms\cite{DBLP:journals/pvldb/GuoLXYYLCXLLCQW22, DBLP:conf/sigmod/WangYGJXLWGLXYY21}.
Existing VDBMS can be roughly categorized into two types: \textit{native} systems and \textit{extended} systems\cite{vector_survey_2023}.
The former, such as Chroma\cite{Chroma} and EuclidesDB\cite{euclidesdb}, are tailored for vector data management, while the latter, such as AnalyticDB-V\cite{DBLP:journals/pvldb/WeiWWLZ0C20}, are built on top of an existing data management system.

In this work, we first introduce a visualization pruning method to filter out less informative visualizations, \jn{which} effectively reduce storage demands and \jn{maintain} the quality of stored visualizations.
We then integrate Chroma\cite{Chroma} into \tool and utilize an efficient graph-based index to store and index the large collection of visualization references. 
An \textit{Approximate k-Nearest Neighbors} (ANN) algorithm is incorporated to retrieve visualizations.
The combination of pruning with indexing allows for quick retrieval and \jn{efficient} handling \jn{of} timely user queries.

\section{Overview}
\label{sec:overview}
\jn{The} visualization-as-representation framework consists of four modules: \emph{visualization alignment}, \emph{visualization referencing}, \emph{visualization pruning}, and \emph{visualization interaction}, as shown in Figure~\ref{fig:pipeline}.

\stitle{Visualization Alignment} (Sect.~\ref{sec:chart_alignment}).
As the cornerstone of our framework, an MLLM is fine-tuned to establish a single joint embedding space for various modalities, including chart ($\mathbb{C}$), text ($\mathbb{T}$), and sketch ($\mathbb{S}$).
To address the lack of suitable training data, we first curate a new dataset with chart-text pairings augmented from existing Chart-to-Text dataset\cite{chart-text}, and chart-sketch pairings through user labeling.
Next, we fine-tune an MLLM on our composite dataset, employing a Transformer\cite{gpt2_2019} for text modality, and ViT\cite{vit_2020} encoders for chart and sketch modalities, respectively.
To achieve alignment across modalities within the joint embedding space, we implement a hinge-based triplet ranking loss and a two-level cross-entropy loss. 
Through this process, we successfully align the three modalities and \jn{achieve} strong performance across four evaluation metrics\jn{, all} surpassing a threshold of 0.85.

\stitle{Visualization Referencing} (Sect.~\ref{ssec:vis_referencing}).
In the \emph{visualization referencing} module, we generate visualization references that comprehensively capture key data patterns inherent in an input table.
This module employs a two-stage approach that first decomposes the input table $D$ into data facets, each representing a subset of the table's temporal or structural characteristics. These facets are then mapped into fixed-size visualization references, ensuring a rich and diverse representation of the data that captures both long- and short-term patterns, trends, anomalies, and temporal relationships.

\stitle{Visualization Pruning} (Sect.~\ref{ssec:vis_pruning}).
In the \emph{visualization pruning} module, we introduce pruning and indexing techniques to enable timely visualization reference retrieval.
Specifically, visualization pruning filters out less informative visualization references by computing the Euclidean distance between visualization vectors and applying a pruning threshold.
This process reduces the number of stored references while preserving essential data change patterns.
Concurrently, visualization indexing leverages an open-source vector database, Chroma\cite{Chroma}, to efficiently store and retrieve visualization references.
Here, we use the fine-tuned MLLM to transform all the pruned visualization references into 512-dimensional vectors, which are then indexed by Chroma for rapid and accurate retrieval.
This combination of pruning and indexing ensures \tool can swiftly manage and retrieve massive visualization references with precision.

\stitle{Visualization Interaction} (Sect.~\ref{ssec:vis_inter}).
For \emph{visualization interaction}, we introduce a ``decomposing-execution-filling'' strategy to enable the connection of multimodal user queries with the stored embeddings of visualization references.
Specifically, the user query is first decomposed into a visualization retrieval task. Second, the retrieval task is executed in Chroma to obtain the retrieved visualization reference. 
Afterward, with the help of the fine-tuned MLLM, the corresponding trend word of the retrieved visualization reference is generated and then filled into the textual response to the user.
This strategy mitigates the stochastic nature of LLMs by decomposing queries into deterministic retrieval operations, reducing hallucinations and ensuring consistent results for identical inputs.
An interactive visual interface is also developed to enable analysts to perform iterative table reasoning tasks, including trend identification, anomaly detection, and pattern comparison. 
\section{Multimodal Visualization Alignment}
\label{sec:chart_alignment}
We \jn{aim} to \jn{build} an MLLM that aligns visualization modalities of chart, sketch, and text \jn{within} a single joint embedding space.
This section presents the data preparation \jn{process}, including data augmentation and user labeling strategies, to address the issue of lacking \jn{a} training dataset (Sect. \ref{ssec:data_pre}).
\jn{Next, we introduce} modality binding for establishing a joint embedding space that \jn{unifies} chart, sketch, and text modalities (Sect. \ref{ssec:align}).
In the end, we present the implementation details (Sect. \ref{ssec:imp}).

\begin{figure}
    \centering
    \includegraphics[width=0.985\linewidth]{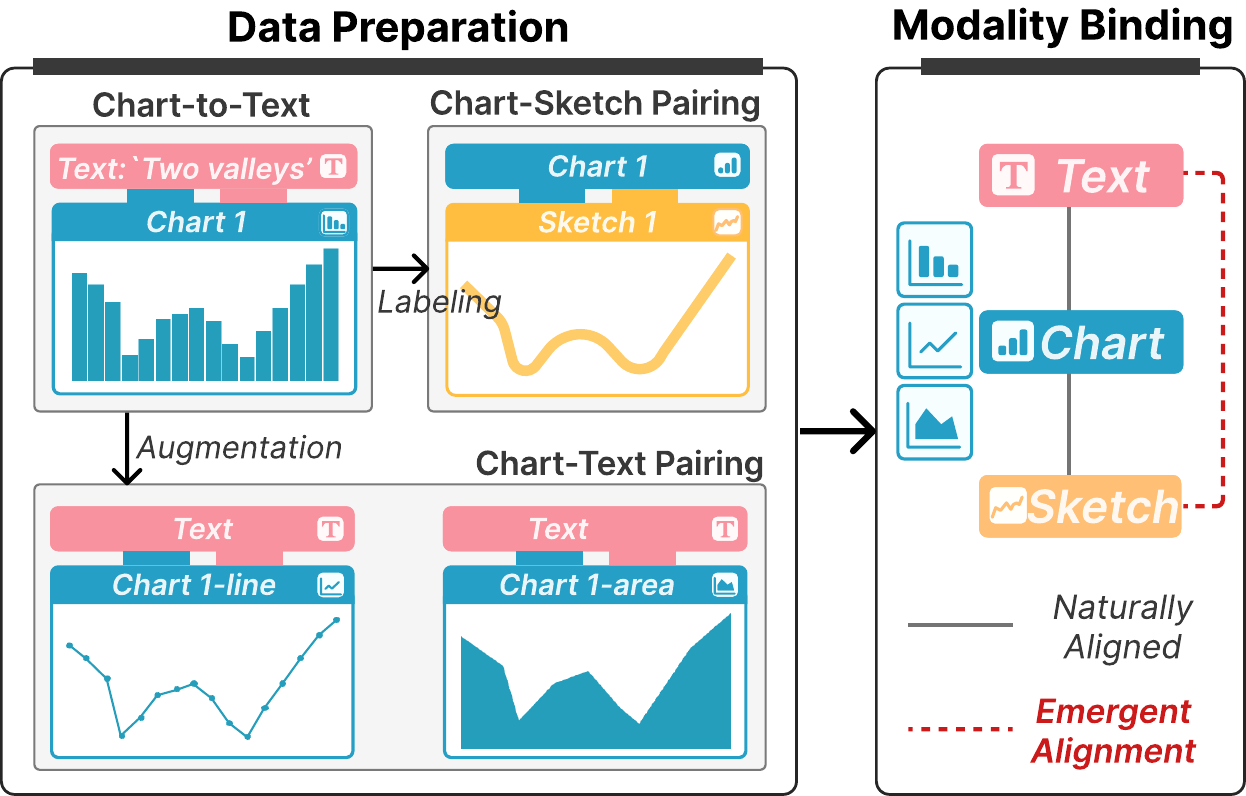}
    \vspace{-2mm}
    \caption{An overview of \emph{visualization alignment} module, which contains data preparation for chart-text and chart-sketch pairings, and \jn{jointly aligns} chart, sketch, and text modalities.}
    \label{fig:alignment}
    \vspace{-2mm}
\end{figure}

\subsection{Data Preparation}\label{ssec:data_pre}

To train the multimodal visualization alignment model, we need to first prepare a dataset containing user intentions, which is not covered by the current open-source datasets.
To achieve the alignment of sketches, charts, and text, we employ two strategies to construct the training data: 1) data augmentation for chart-text pairing, and 2) user labeling for chart-sketch pairing, as illustrated in Figure~\ref{fig:alignment}.

\stab(1) \textbf{Data augmentation for chart-text pairing.}
To \jn{compensate} for the lack of data change pattern recognition in existing table reasoning works, we focus on charts that describe data change patterns.
We first survey existing chart-text datasets, and 
select the Chart-to-Text dataset\cite{chart-text} as the original dataset, since it offers textual descriptions and underlying data tables for charts.
Since we focus on data change patterns \jn{present} in charts, we filter trend words from the textual descriptions.
Initially, we manually extracted the contained trend words in five textual descriptions.
These five descriptions with their corresponding trend words served as prompts for GPT-4 to perform an accurate trend word filtering process on the large Chart-to-Text dataset.
For example, the description for chart 110 is \q{Share of voters casting ballots by mail has steadily risen since 1996}, with the corresponding trend word `rise'. 
The description for chart 438 is \q{In Turkey, daily Arabic internet searches for Greece peaked \jn{shortly after midnight}}, with the corresponding trend word `peak'.
In the filtering process, we find diversity in both trend words and charts.
For trend words, we notice that different trend words may indicate the same change pattern, such as `increasing', `upward', and `rising', which typically indicate an upward trend in the data.
We construct a dictionary of similar trend words \jn{by} grouping words \jn{that describe} similar patterns into the same category, \jn{resulting in} 23 categories ($N_T = 23$).
For charts, we find that three chart types: line, bar, and area charts, are widely chosen to describe the data change patterns.
To balance the number of three chart types contained in the final chart-text pairing dataset, we supplement each filtered chart with two other types of charts.
In total, we create 5,010 chart-text pairings.

\stab (2) \textbf{User labeling for chart-sketch pairing.}
Existing public hand-drawn sketch datasets do not adequately meet our need to convey users' intentions about data change patterns.
Specifically, existing sketch datasets are not suitable for our work due to the following points.
1) Most open-source large human sketch datasets (\eg,\cite{humansketch_2012}) focus on how users draw objects, rather than on visualizations. 
2) LADV\cite{ladv_2020} and VISAtlas\cite{visatlas_2022} are designed for chart type matching based on pixel-wise similarity, but not the data change patterns.
3) Sketch query systems, such as Qetch\cite{qetch_2018}, focus on data-level similarities between hand-drawn sketches and the raw data, lacking correspondence with fixed-size charts.

In order to collect chart-sketch pairings that conform to users' intentions about data change patterns, we \jn{invited} users to draw sketches and evaluate chart-sketch pairings.
1) \emph{Setup.} Every user \jn{was} given a training session to get familiar with experimental procedures and hand-drawn board operations.
2) \emph{Reference chart selection.} Every user \jn{was} given 100 real charts selected from our chart-text pairing dataset. The selected charts cover 23 categories of trend words, and contain bar, line, and area charts.
3) \emph{Drawing.} The selected charts \jn{were} displayed to the user one by one, and the user \jn{drew} a sketch \jn{representing} the change patterns in the displayed chart on \jn{a} free-scale board.
4) \emph{Evaluation/Redraw.} After drawing all the given charts, we \jn{showed} \jn{the user} each sketch with the corresponding chart one by one, and the user could delete or redraw the sketches they \jn{were} not satisfied with.
A total of 10 users (six women, four men) aged from 18 to 30 were invited.
They all \jn{had} knowledge of data visualization and \jn{were experienced} in drawing sketches.
In the end, we \jn{filtered} low-quality sketches, defined as those with unclear lines and incomplete patterns, resulting in 1,214 pairs of chart-sketch pairings ready for training.

The final training dataset, including all chart-text and chart-sketch pairings, has been released at \url{https://github.com/HKUST-CIVAL/VisTR}. 

\begin{figure}[t]
    \centering
    \includegraphics[width=0.99\linewidth]{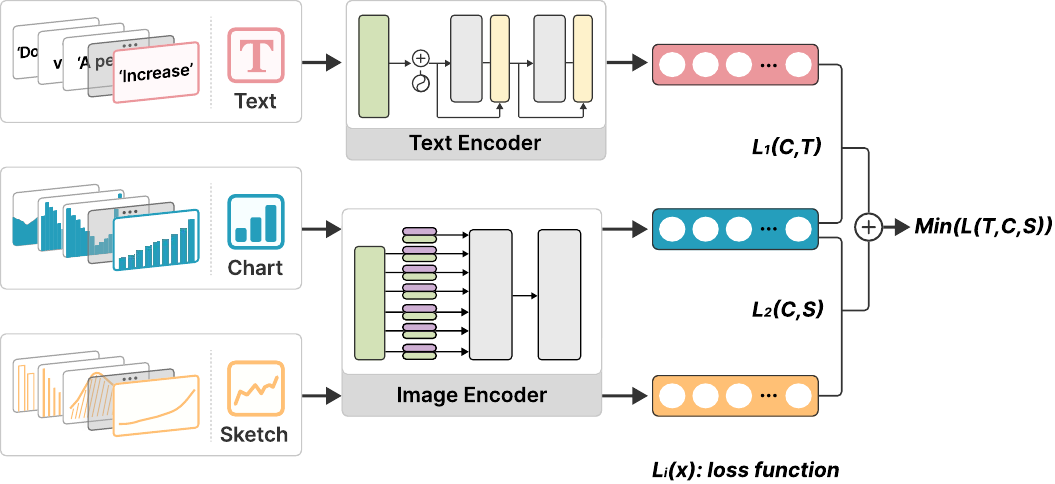}
    \vspace{-4mm}
    \caption{The alignment pipeline for three modalities. A shared ViT image encoder processes both chart and sketch modalities, while text uses a separate Transformer encoder. The chart modality is set as the supervisor to guide the alignment process.}
    \vspace{-4mm}
    \label{fig:space}
\end{figure}

\subsection{Modality Binding}\label{ssec:align}
We aim to bind all text $\mathbb{T}$, chart $\mathbb{C}$, and sketch $\mathbb{S}$ modalities together \jn{within} a joint embedding space.
As shown in Figure~\ref{fig:space}, we employ a two-stage strategy.
First, we utilize modality-specific encoders to transform each modality into embeddings of uniform length\cite{structchart_2023}. 
All three modalities are processed using the Transformer architecture, leveraging its self-attention mechanism to effectively capture complex dependencies within each modality.

Inspired by studies on sketch-based image retrieval tasks\cite{Lyou_2024_WACV}, we treat sketches and charts as distinct modalities due to their differing characteristics, even though they can both be embedded via the same image encoder.
Given a chart $C_i$ with its corresponding observation $T_i$ in text modality, and observation $S_i$ in sketch modality, we encode them into normalized embeddings:
$p_i = f(C_i)$, $q_i = f(S_i)$, and $k_i = f(T_i)$, where $f$ is a 63M-parameter 12-layer 512-wide Transformer with the architecture modifications with 8 attention heads\cite{gpt2_2019} for text modality, and a Vision Transformer (ViT)\cite{vit_2020} with a patch size of 32 for chart and sketch modalities.
We add a modality-specific linear projection head on each encoder to obtain a fixed-size $d$-dimensional embedding, which is normalized, and set as $p_i$, $q_i$, and $k_i$ for loss calculation.
Besides facilitating learning, this setup allows us to initialize a subset of the encoders with weights from pre-trained CLIP\cite{CLIP_2021} models.
This initialization not only provides a strong starting point for our MLLM but also enables it to better understand and align the semantic content across different modalities.

During the second stage, we 
elect the chart modality, $\mathbb{C}$, as the supervised modality because visualization references are in the form of charts, and the chart-text and chart-sketch pairings that span a wide range of change patterns are well collected.
Inspired by VSE++\cite{vse_2017}, we introduce a two-level cross-entropy and a bidirectional hinge-based triplet ranking loss, which makes the matched pairs have \jn{lower} entropies than unmatched ones.
The formulation for the two cross-entropy functions is specified as:
\begin{equation}
    H_1(C, T) = - \sum (p_i \times \log(k_i)),
\end{equation}
\begin{equation}
H_2(C, S) = - \sum (p_i \times \log(q_i)),
\end{equation}
where $H_1$ calculates the difference between charts $C$ and text $T$, and $(C, T)$ is the matched positive chart-text pair. 
$H_2$ calculates the difference between charts $C$ and sketches $S$, and $(C, S)$ is the matched positive chart-sketch pair.

We consider the loss function for the chart-text pairings as $L_1 (C, T)$, and the loss function for the chart-sketch pairings as $L_2 (C, S)$.
Take chart and sketch modalities as an example, the loss is as follows:
\begin{align}
L_2 (C, S) = [\alpha + H_2(C', S) - H_2(C, S)]_{+} \nonumber \\
+ [\alpha + H_2(C, S') - H_2(C, S)]_{+},
\end{align}
where $\alpha$ denotes the margin parameter, and $[x]_{+} \equiv \max(x,0)$.
$C' = \arg \max_{X \neq C} H_2(X, S)$ and $S' = \arg \max_{Y \neq S} H_2(C, Y)$ denote the hardest negative corresponding to the positive pair $(C, S)$.
The loss makes the chart embedding $p_i$ and the sketch embedding $q_i$ closer in the joint embedding space, and thus aligns $\mathbb{C}$ and $\mathbb{S}$.
Set $L$ as the final loss of three modalities in the embedding space, that is, the sum of the two ranking losses:
$L(T, C, S) = L_1(C, T) + L_2(C, S).$
By minimizing both the loss between chart-text modalities and the loss between chart-sketch modalities, we align the embeddings of the three modalities \jn{within} a joint embedding space.

\subsection{Implementation Details}\label{ssec:imp}
We take a pre-trained CLIP text-image model and fine-tune it on the selected chart-text and chart-sketch pairings.
In the training phases, we employ a strategy where we \jn{freeze} the ViT encoder and conduct fine-tuning for 30 epochs with a batch size of 32, and a learning rate of $1e^{-5}$.
The model parameters \jn{are} optimized using the AdamW optimizer, employing Betas $(0.9, 0.98)$, weight decay of 0.001, combined with a StepLR learning rate schedule. 
This static scheduler ensures that the learning rate gradually decreases during fine-tuning, facilitating convergence towards a local optimum. 
\jn{By using} the modalities paired with charts, \textit{i.e.}, pairs of $(\mathbb{C},\mathbb{T})$ and $(\mathbb{C},\mathbb{S})$, we align each of the embeddings from text modality $\mathbb{T}$ and sketch modality $\mathbb{S}$ to those from charts.
The pairs \jn{$(\mathbb{T},\mathbb{S})$} are aligned even though we only train using the pairs \jn{$(\mathbb{C},\mathbb{T})$} and \jn{$(\mathbb{C},\mathbb{S})$}.
We refer to the statement that $(\mathbb{C},\mathbb{S})$ and $(\mathbb{C},\mathbb{T})$ are naturally aligned, and the text $\mathbb{T}$ and sketch modalities $\mathbb{S}$ constitute an \jn{emergent} alignment.

\section{\tool System}
With the MLLM for multimodal visualization alignment, we build \tool system that empowers data analysts and researchers to intuitively interact with time-series tabular data.
This section details the implementation of \emph{visualization referencing} (Sect. \ref{ssec:vis_referencing}), \emph{visualization pruning} (Sect. \ref{ssec:vis_pruning}), and \emph{visualization interaction} (Sect. \ref{ssec:vis_inter}) in \tool system.

\subsection{Visualization Referencing}\label{ssec:vis_referencing}
This module is designed to ensure that the generated visualization references can effectively capture comprehensive data patterns inherent in the raw tabular data.
To this end, we design a two-stage approach that first generates data facets to capture all underlying data change patterns and then maps all data facets into visualization references.

\stitle{(1) Facet Generation.}
In the first stage, we aim to capture meaningful data facets of the input table as comprehensively as possible.
A naive exhaustive enumeration would generate all possible data facets, producing too many candidates, many of which lack meaningful data trends.
To avoid the inefficient data facet generation, we adopt the idea of subset embedding~\cite{sen_2021}, indicating that patterns of multi-dimensional data are often related to subsets rather than records.
Specifically, we adopt a \jn{multifaceted} process that incorporates a dual strategy of smoothing and segmentation.

Consider an input time-series table $D(ts, d_1, d_2, \cdots, d_n)$, where each $d_i$ represents a distinct variable containing a sequence of data points indexed by the timestamp ($ts$).
The first step is to smooth each variable $d_i$ to retain essential patterns while filtering out the noise.
To achieve this, we employ Gaussian Smoothing, a technique favored for maintaining shape consistency under the Fourier transform\cite{smoothing_2012}.
Formally, for each variable $d_i$, the smoothed series $\widetilde{d_i}$ is obtained as: $\widetilde{d_i} = GaussianSmoothing(d_i)$.
Compared with widely used Moving Average (MA) or Weighted Moving Average (WMA)\cite{timetuner}, Gaussian Smoothing can preserve key features during smoothing and provide flexibility in parameter control.

After smoothing, we segment each smoothed variable $\widetilde{d_i}$ into distinct facets to extract meaningful and reliable patterns.
The segmentation process is defined as $R(\widetilde{d_i}) = \{r_1(\widetilde{d_i}), r_2(\widetilde{d_i}), \cdots, r_m(\widetilde{d_i})\}$, where $r_j(\widetilde{d_i})$ represents a subset (termed a facet) of $\widetilde{d_i}$.
To ensure each data facet reflects a valid data pattern, we incorporate the Page-Hinckley Test (PHT)\cite{PHT_2004} to detect critical points where shifts occur, thereby delineating the boundaries between different facets.
Specifically, the segmentation process can be formalized as $R(\widetilde{d_i}) = PHT(\widetilde{d_i})$.
Additionally, we aggregate continuous subsets to form new facets, enabling the inclusion of segments with varying lengths.

\stitle{(2) Visual Mapping.}
The second stage is visual mapping, where we transform each data facet into a standard visual representation.
This process leverages the \emph{matplotlib} library to generate fixed-size visualization references \jn{without} axes, specifically utilizing bar, line, and area charts to \jn{represent} the trend within each data facet.
All references are rendered at a fixed 16:9 aspect ratio (\eg, $512 \times 288$ pixels) to maintain shape consistency across different display contexts.
The uniformity of fixed-size visualization references improves the fine-tuned MLLM's capacity to analyze and process them.
For each data facet $r_j(\widetilde{d_i}) \in R(\widetilde{d_i})$, the visual mapping process transforms $r_j(\widetilde{d_i})$ into a fixed-size visualization reference $c_j$, which is formalized as $c_j = vmap(r_j(\widetilde{d_i}))$.
Specifically, $vmap$ converts every facet into three chart types (bar, line, and area) to maximize cross-modal alignment with diverse user inputs.
The collection of all visualization references derived from the data facets of a particular variable $d_i$ is represented as $C(d_i) = \{c_1, c_2, \cdots, c_m\}$.
Collectively, for all variables in the input table $D$, the visualization references are denoted as $C = \bigcup_{i=1}^nC(d_i) = \{c_{i,j} | d_i \in D, r_j(d_i) \in R(d_i)\}$.

\subsection{Visualization Pruning and Indexing} \label{ssec:vis_pruning}
\begin{figure}[t]
    \centering
    \includegraphics[width=0.95\linewidth]{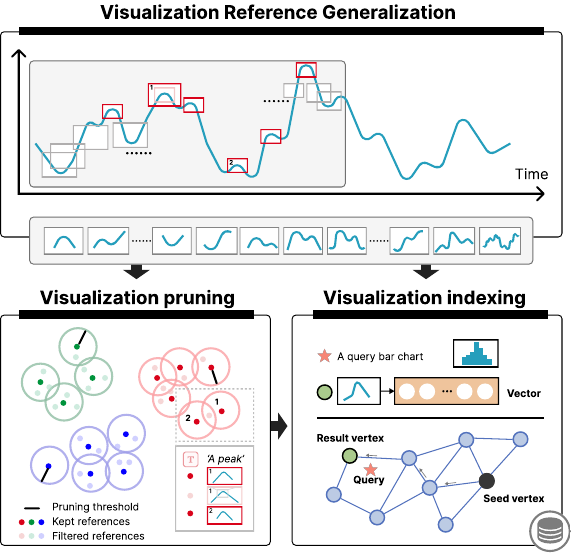}
    \vspace{-3mm}
    \caption{The pipeline of \emph{visualization referencing} (top) and \emph{visualization pruning and indexing} (bottom) modules, which are used to accelerate the storage and retrieval process.}
    \label{fig:filter}
    \vspace{-5mm}
\end{figure}

For a given table, we generate a vast collection of visualization references, each of which is subsequently mapped into a vector representation (\ie, an embedding).
The large number of visualization references poses significant challenges to efficient storage and retrieval.
To address these challenges, we design a combined approach of visualization pruning and indexing, as illustrated in Figure~\ref{fig:filter}.

\stitle{Visualization Pruning.} 
Here, the key idea is to filter out less informative visualization references in advance while retaining those that provide significant insights.
For all visualization references for a variable $d_i$, \jn{we may encounter} a scenario where two references exhibit similar change patterns, and one is entirely encompassed by the other.
To quantify the similarity between visualization vectors, we employ Euclidean distance $d(c_j,c_k)$ between the two vectors $c_j$ and $c_k$. 
This choice is preferable as it measures the spatial differences between vectors, which intuitively correspond to visual similarities in visualizations, enabling robust detection of patterns without requiring normalization adjustments.
We introduce a pruning threshold to determine whether two visualization vectors are sufficiently similar to warrant pruning.
If $d(c_j,c_k)$ is less than the pruning threshold, the vector covering a shorter time range is filtered out.
An example scenario is illustrated in Figure~\ref{fig:filter}, where the `a peak' pattern exists in pink-boxed visualization references.
In this case, the light-pink-boxed one has less information and is consequently filtered out.

The pruning threshold acts as a predefined parameter that affects the number of stored visualization references.
To assess its impact, we designed an experiment with four values: 0.5, 0.8, 1, and 1.5, and applied the thresholds to tables of varying row counts.
Specifically, the row numbers were considered based on the typical sizes used in existing LLM-based table reasoning methods, which typically support data tables with fewer than 50 rows.
For instance, TabFact\cite{tabfact_2020}, a commonly used dataset containing over 16k Wiki tables, has an average of 12.96 rows per table (Max.=48, Min.=1, SD=8.46).
Given the focus on time-series data tables and long-term trend analysis, we extended the experiment to tables with up to 1000 rows, approximately encompassing four trading years.
We calculated the average number of visualization references across three sub-tables with corresponding table sizes extracted from the DJI. data table for each pruning threshold.

\begin{figure}[t]
    \centering
    \includegraphics[width=0.95\linewidth]{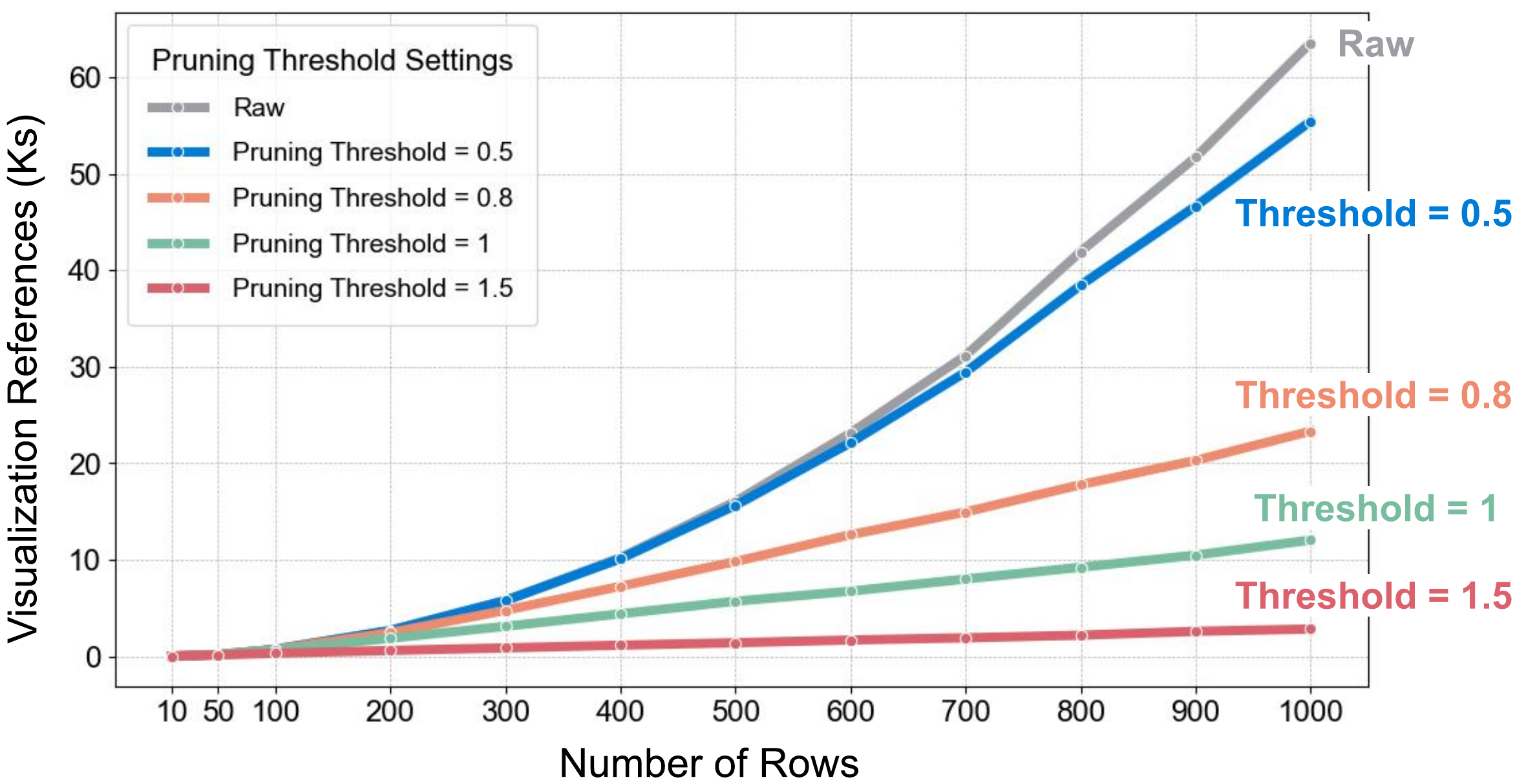}
    \vspace{-3mm}
    \caption{Relationship between the number of visualization references, table rows, and the pruning threshold. The remaining visualization references are close to 20\% of the raw number when the threshold is set to 1.
    }
    \vspace{-3mm}
    \label{fig:threshold}
\end{figure}

Figure~\ref{fig:threshold} shows the relationship between the number of stored visualization references, the number of data table rows, and the pruning thresholds.
When the pruning threshold is set to 1, the number of stored visualization references approximates 20\% of the raw reference number.
This aligns with the Pareto principle (80/20 rule)\cite{powerlaw_2005}, which is widely leveraged in various optimization strategies to balance efficiency and information retention.
As such, we set the pruning threshold to be 1, which strikes a balance between maintaining a rich representation set and reducing the storage size.

\stitle{Visualization Indexing.}
For efficient retrieval of visualization references, we utilize an open-source vector database named Chroma\cite{Chroma}.
Specifically, we first apply our fine-tuned MLLM to encode all visualization references into 512-dimensional vectors. 
Chroma automatically maintains an index document when encoding all visualization references for retrieval.
These indices compress vectors into \jn{fixed-size} codes \jn{and store} them in an array. 
For similarity measurement, we employ the dot product function, as it efficiently computes the directional alignment between vectors in high-dimensional space.
By leveraging Chroma's advanced indexing capabilities, \tool efficiently speeds up the retrieval process to handle vast amounts of visualizations, ensuring both speed and accuracy.

\begin{figure*}[htbp]
    \centering
    \includegraphics[width=0.99\textwidth]{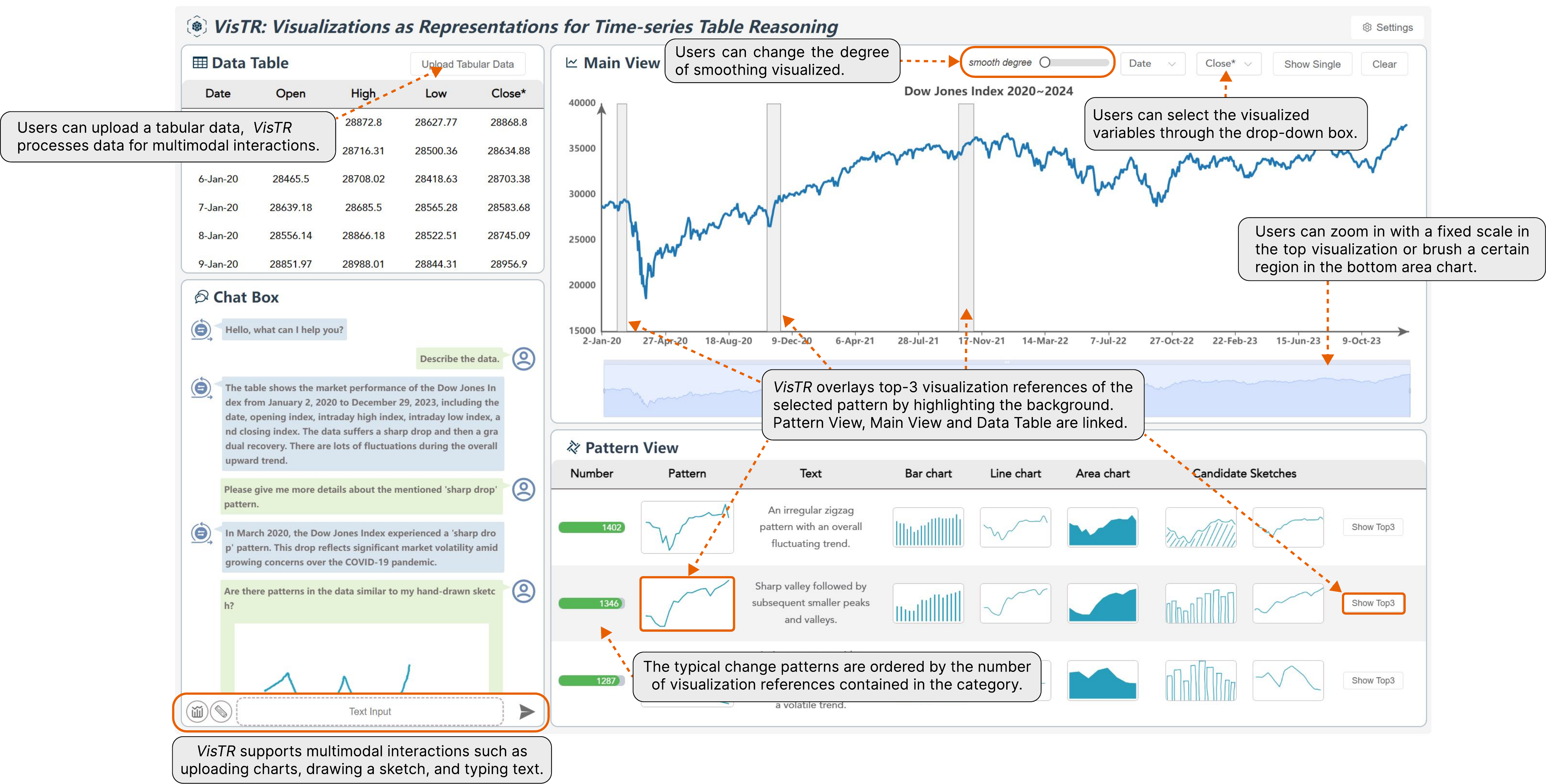} 
    \vspace{-3mm}
    \caption{The interactive interface of \tool, which contains four main views and supports multimodal interactions for time-series table reasoning.}
    \vspace{-3mm}
    \label{fig:interface}
\end{figure*} 

\subsection{Visualization Interaction}  \label{ssec:vis_inter}

\stitle{The Decomposing-Execution-Filling Strategy.}
Inspired by Dater\cite{dater_2023}, we introduce a `decomposing-execution-filling' strategy to decompose and parse multimodal user queries and return results that meet user expectations.
Specifically, we first decompose a user's table reasoning queries and convert them into visualization reference retrieval operations, similar to text-to-SQL parsing (\eg,\cite{parsingsurvey_2022, proton_2022}). 
Afterward, we execute the visualization retrieval operations in Chroma and get the result for back-filling the final response.

Here, we employ two LLMs with in-context learning using prompts to decompose user queries and complete outputs, respectively.
Below we illustrate an example using the case shown in Figure~\ref{fig:teaser}-Q1.
Given the query ``\textcolor{grey}{What is the price trend of Apple during March?}'', we first decompose the question into a retrieval task ``\textcolor{teal}{return the visualization reference of Apple price in March}'', and a cross-modal alignment task ``\textcolor{teal}{recognize the trend in the retrieved visualization reference}''.
The corresponding visualization reference is first retrieved from Chroma, and then aligned with the `two peaks' pattern to generate the final textual response ``\textcolor{teal}{There is a two-peak price trend in Apple during March}''.
This strategy enhances reliability by mitigating the stochastic nature of LLMs and reducing hallucinations\cite{hallucination_2023}, as core results stem from retrieval rather than probabilistic generation.

\stitle{Visual Interface.}
In addition to textual responses, \tool incorporates a user-friendly interface to support intuitive table reasoning and exploration.
As shown in Figure~\ref{fig:interface}, the interface includes four main views: (A) {\tt Data Table}, (B) {\tt Main View}, (C) {\tt Chat Box}, and (D) {\tt Pattern View}.
When a user uploads a table, the back-end of \tool processes the tabular data into visualization references and stores them in the vector database, while the front-end interface displays the uploaded table in {\tt Data Table}, and presents the visualization for the data in {\tt Main View}.
To maintain consistency with the fixed-size visualization references (generated at a standard 16:9 aspect ratio with $512\times288$ pixels), {\tt Main View} adopts a 32:9 aspect ratio for the chart display area.
This wider ratio facilitates effective contextual visualization in long time series. 
The individual selected 16:9 visualization reference is displayed in the central portion of this view to conform to common widescreen displays.
This uniform aspect ratio prevents shape distortion due to varying display proportions and ensures that the trends in user-drawn sketches or uploaded charts match the shapes of the rendered charts.
By default, {\tt Main View} uses an area graph for the overall context information and a line chart for the details.
Various operations, such as selecting different variables or combinations of variables, are enabled in the drop-down box in the upper right corner of {\tt Main View}.
For the display of multiple variables, we use a multi-line chart with colors to encode different variables.
The bottom area chart also allows for fixed-scale zooming and brushing, allowing users to observe detailed variable changes.

{\tt Chat Box} provides a dialogue window that supports multi-round conversations and multimodal inputs, including uploading charts, drawing sketches, and typing text.
Textual responses to user queries are also displayed in {\tt Chat Box}.
The responses are also connected to {\tt Main View}, where the corresponding visualization reference is highlighted.
{\tt Pattern View} presents groups of typical data change patterns for the variable selected in {\tt Main View}.
These groups are ordered from top to bottom, with the most populated data change pattern at the top and the least at the bottom.
Users can select a specific group of data change patterns, and then the top-3 visualization references of this pattern will be \jn{overlaid} by highlighting the background.
If there is an overlap among the displayed visualization references, the overlapping area \jn{is rendered with} a darker background color.
Users can click the `show' button located in the upper right corner of {\tt Main View} to achieve a circular review of each one.
Additionally, {\tt Pattern View} provides multimodal information (description text, different chart types, and candidate sketches) for a single variable's typical change patterns, enabling analysts to quickly identify and understand them.

\section{Evaluation}\label{sec:evaluation}

To thoroughly evaluate \tool, we conduct a quantitative evaluation that assesses the alignment ability of our fine-tuned MLLM (Sect.~\ref{ssec:evaluation}).
Next, we present two representative usage scenarios (Sects. \ref{ssec:study1} \& \ref{ssec:study2}), to illustrate how \tool can support time-series table reasoning tasks through multimodal interactions.
Last, we provide qualitative user feedback on the \tool system (Sect.~\ref{ssec:interface_feedback}).

\subsection{Effectiveness of Multimodal Visualization Alignment}\label{ssec:evaluation}

To rigorously assess the performance of \tool for multimodal visualization alignment, we perform a quantitative evaluation across three aspects: 
1) \emph{text-chart retrieval} measures the underlying CLIP model's ability to retrieve relevant chart images from textual queries;
2) \emph{sketch-TS query} evaluates the effectiveness of \tool in identifying correct time-series data from sketch-based inputs;
and 3) \emph{TS-text reasoning} examines the model's ability to perform text reasoning when provided with raw time-series data.

\subsubsection{Text-Chart Retrieval}
To faithfully evaluate \tool's ability to align chart-text pairs, we ensure that the evaluation dataset includes chart types of bar, line, and area charts, as well as all $T_N=23$ categories of trend words.
We randomly select 20\% of the data for each chart type and each trend category from the chart-text pairing dataset for evaluation.

\stitle{Baselines}.
Our MLLM is a fine-tuned CLIP model on the composite dataset.
To highlight the necessity and impact of fine-tuning, we compare it with the zero-shot CLIP and OpenCLIP\cite{openclip_2021} models as an ablation study.
Additionally, we validate the effectiveness of our model by comparing it against \jn{fine-tuned} OpenCLIP\cite{openclip_2021} \jn{using the same dataset as ours}, \jn{to confirm} the appropriateness of our model selection.

\stitle{Metrics}.
Considering the imbalance among different trend word categories, we set two metrics for the evaluation:
\begin{itemize}
\item
Top-1 Accuracy (Acc). The accuracy calculates the ratio of \jn{accurately} identified pairs to the total number of pairs examined.

\item
Weighted F1-Score (WF). WF adopts the weighted average considering the unevenness of the categories, which is calculated as: 
$WF = \frac{1}{T_N} \sum_{i=1}^{T_N} \frac{2 \times P_i \times R_i}{(P_i+R_i)}$,
where $R_i$ and $P_i$ are the recall and precision of the $i^{th}$ trend-word category, respectively.
\end{itemize}

\begin{table}\footnotesize
\centering
\caption{Experimental results on text-to-chart alignment test for all and individual chart types.}
\label{table:eval}
\renewcommand{\arraystretch}{1.1}
\begin{tabular}{cccccc}
\hline
\rowcolor[gray]{.9}
 Model &      & All & Bar & Line & Area \\ \hline

\multirow{2}{*}{\makecell[c]{Ours\\(Fine-tuned CLIP)}} & Acc (\%)  & \textbf{89.02}  & \textbf{86.29} &  \textbf{89.00}  & \textbf{89.12}\\
    & WF (\%)  & \textbf{87.45}  & \textbf{85.91} & \textbf{86.54} & \textbf{87.79} \\ \hline       
\multirow{2}{*}{\makecell[c]{Fine-tuned\\OpenCLIP}} & Acc (\%)  &  50.74 & 50.36 & 51.86 & 50.24 \\
    & WF (\%)  &  49.77 & 48.04 & 46.41 & 49.73  \\ \hline  
\multirow{2}{*}{\makecell[c]{Zero-shot\\CLIP}} & Acc (\%)  & 0.82  & 1.10 & 0.72 & 1.02 \\
    & WF (\%)  & 0.36 & 0.29 & 0.11 & 0.07 \\ \hline  
\multirow{2}{*}{\makecell[c]{Zero-shot\\OpenCLIP}} & Acc (\%) & 7.71 & 8.20  & 5.27 & 1.08 \\
    & WF (\%)  & 5.37 & 3.78 & 1.95 & 0.30 \\ \hline  
\end{tabular}
    \vspace{-3mm}
\end{table}

\stitle{Results}. 
The results are presented in Table~\ref{table:eval}.
Our fine-tuned MLLM demonstrates robust performance in aligning chart and text modalities, with overall accuracy and weighted F1-score exceeding 85\% across all chart types.
The results for each individual chart type also provide substantial evidence of the model's effectiveness. 
However, bar charts yield slightly lower metrics than other chart types.
This discrepancy may be attributed to the inherent characteristics of bar charts, which utilize separated bars to encode data, resulting in less distinctiveness compared to the continuous line marks used in line charts and the filled area marks used in area charts.

The comparisons between fine-tuned CLIP and zero-shot CLIP models confirm that fine-tuning is crucial for enhancing model performance in the specific domain.
While zero-shot OpenCLIP slightly outperforms zero-shot CLIP, fine-tuned CLIP significantly surpasses fine-tuned OpenCLIP.
Zero-shot OpenCLIP outperforms zero-shot CLIP, \jn{which is} probably due to more extensive pre-training data that includes more diverse chart images.
However, after fine-tuning, CLIP's \jn{architecture is} better suited for our requirements.
Its superior results validate our choice of CLIP as the base model.

\begin{table}[t]
\centering
\caption{Quantitative comparison of \tool with Qetch and existing MLLMs in sketch-TS query and TS-text reasoning tasks. Best results are in bold.}
\label{table:quan}
\footnotesize 
\renewcommand{\arraystretch}{1.1}
\resizebox{\columnwidth}{!}{
\setlength{\tabcolsep}{2pt}
\begin{tabular}{l|cccc|cc}
\toprule
\multirow{2}{*}{\textbf{Method}} & \multicolumn{4}{c|}{\textbf{Sketch-Ts Query@3}} & \multicolumn{2}{c}{\textbf{Ts-Text Reasoning}} \\ \cmidrule(lr){2-5} \cmidrule(lr){6-7}
& DTW$\downarrow$ & PCC$\uparrow$ & Human$\uparrow$ & Lat.$\downarrow$ & LLM Judge$\uparrow$ & Lat.$\downarrow$ \\ \midrule
VisTR & \textbf{2.09} & 0.46 & 0.77 & 150.06 ms  & \textbf{0.77} & \textbf{326.83 ms} \\
\hline
Qetch (n=20) & 4.53 & 0.48 & \textbf{0.81} & 16.67 ms & \multirow{2}{*}{\xmark} & \multirow{2}{*}{\xmark} \\ 
Qetch (n=40) & 5.31 & \textbf{0.52} & 0.75 & \textbf{11.28 ms} &  &  \\ \hline
\multicolumn{7}{l}{\textit{Multimodal LLMs}} \\
Claude-4-Sonnet & 5.92 & 0.40 & 0.49 & 5.56 s & 0.72 & 4.83 s \\
Gemini-2.5-Pro & 3.50  & 0.50 & 0.57 & 39.02 s & 0.28 & 4.60 s \\
GLM-4.5v  & 3.65 & 0.12 & 0.36 & 30.32 s & 0.62 & 4.59 s \\ \bottomrule
\end{tabular}}
\end{table}


\vspace{1.5mm}
\subsubsection{Sketch-TS Query}
Next, we evaluate \tool in sketch-to-TS query task, where the goal is to locate the most similar subsequence within a long time series given a hand-drawn sketch.
This is a representative task in exploratory scenarios where analysts express vague or imprecise patterns visually.
The used dataset measures the daily price movements of 30 large US companies in the Dow Jones Index (DJI.), spanning from January 2nd, 2020 to December 29th, 2023. 
The data includes five columns: Date, Open, High, Low, and Close, detailing the trading metrics for each business day.
We randomly generate 100 sketches and evaluate all methods on the DJI. time-series data.

\stitle{Baselines.} 
We compare \tool with both classical sketch-based time-series matching methods and recent MLLMs.
We select Qetch~\cite{qetch_2018}, as it is a representative sketch-based matching algorithm without specifying length or amplitude. 
Following the original Qetch design, the number of sampled points used to approximate sketches and time series is an empirical parameter that depends on the characteristics of time-series data.
To ensure a robust comparison, we evaluate two configurations with 20 and 40 sampled points.

To examine whether general-purpose MLLMs can directly solve this task, we also evaluate three MLLMs with closely aligned release times.
These include two closed-source models (Claude-4-Sonnet and Gemini-2.5-Pro) and an open-source model (GLM-4.5v).
For all MLLMs, both the hand-drawn sketch and the full time series are provided as input, and the models are instructed to return the top-3 most similar subsequences.
All models are accessed via APIs and evaluated under a unified network environment.
All methods operate on the same Gaussian-smoothed time series, ensuring that differences in performance arise from the retrieval and reasoning mechanisms rather than data processing.

\stitle{Metrics}.
We employ both similarity metrics and human judgments to capture complementary aspects of time-series similarity.
Specifically, we report DTW distance and Pearson correlation coefficient (PCC) to quantify similarity between the retrieved subsequences and input sketches, with the same group-average resampling and normalization procedures as in Qetch.
Besides, we collect similarity ratings (on a scale of 0 to 1) from two independent evaluators who were blinded to the model source of each retrieval result.
All similarity metrics and human ratings are computed for the top-3 retrieved results ($@3$).
We also report the average end-to-end latency (Lat.) for each query method.

\stitle{Results.}
Table~\ref{table:quan} presents the comparison results.
Overall, \tool achieves pattern similarity and human ratings comparable to Qetch for sketch-based retrieval.
However, Qetch requires users to have the domain expertise necessary to tune its parameters effectively, while \tool can operate without this manual configuration step.
\jn{Compared to} existing MLLMs, \tool achieves superior performance, especially in human evaluation and latency.
Directly applying general-purpose MLLMs to sketch-to-TS queries presents significant challenges in both reliability and efficiency.
The high latency of MLLMs is prohibitive for interactive systems, and their fixed context windows are often exceeded by typical time-series data (e.g., an average length of 72,317 tokens in this evaluation).
Although some optimization is possible, these fundamental limitations remain. 
Our results instead indicate that leveraging visual representations provides a more effective foundation for sketch-driven queries, enabling a favorable balance between retrieval accuracy, perceptual alignment, and interactive performance.

\vspace{1.5mm}
\subsubsection{TS-Text Reasoning}
We further evaluate \tool's TS-text reasoning capability to generate a textual summary from an input time series. 
This task requires producing a reasonable natural-language description that captures both global trends and local changes, aligning with the multi-level description (\eg, L2/L3-style summaries advocated in VisText~\cite{Vistext_2023}).
We utilize the TSFragment-600K dataset~\cite{t2s_2025}, which contains paired time series and textual summaries of temporal patterns.
We randomly sample 50 instances per time-series length from \{24, 48, 96\}, resulting in 150 test samples in total.

\stitle{Baselines.}
For TS-text reasoning task, conventional sketch-based query methods like Qetch are not applicable. 
Here, we only compare \tool against existing MLLMs used in the previous experiment.

\stitle{Metrics.}
For evaluation, we follow recent evaluation conventions~\cite{geval_2023, judgebench_2024} and employ GPT-4o as an LLM judge to assess the quality of generated descriptions by comparing them with the ground-truth texts.
The judge evaluates each output along four dimensions, including overall trend recognition, local change capture, semantic alignment of trend-related terms, and precision (\ie, avoiding hallucinated trends not present in the input time series).
For each dimension, GPT-4o produces a score ranging from 0 to 1, and the final LLM-as-Judge score is obtained by averaging the scores across all dimensions.
Besides, we report the average end-to-end latency to evaluate the practical usability in interactive analysis scenarios.

\stitle{Results.} 
The right part of Table~\ref{table:quan} presents the comparative results.
\tool demonstrates strong overall performance, achieving the highest LLM-as-Judge score while operating over 10$\times$ faster than general MLLMs (which require 4.5–4.8 seconds per query).
Notably, Gemini-2.5-Pro scores significantly lower, a point that merits further investigation.
\tool's combination of high accuracy and low latency confirms its strong suitability for interactive analysis.

\subsection{Usage Scenario 1: Financial Data Exploration}\label{ssec:study1}
In this usage scenario, we illustrate how a hypothetical market analyst, Mike, explores financial time-series data using \tool.
Mike first loads the DJI. data as used in sketch-TS query task into \tool and selects `Close' as the displayed variable in {\tt Main View}. 
\tool initially generates 64,261 visualization references, among which 13,823 visualizations are pruned for exploration.

\begin{figure}[t]
    \centering
    \includegraphics[width=0.94\linewidth]{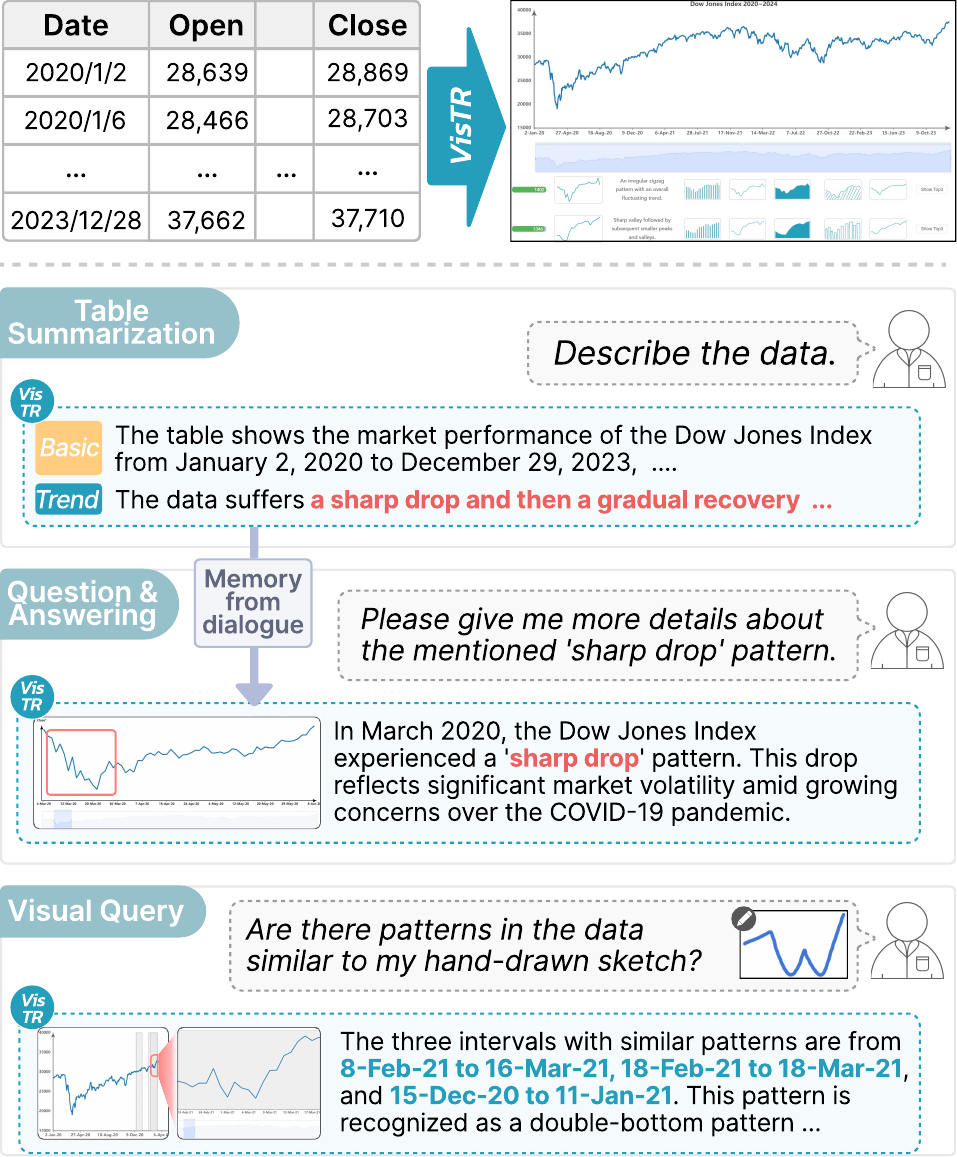}
    \vspace{-2mm}
    \caption{Usage Scenario 1: data exploration for DJI. from 2020 to 2023. Mike gets a complete understanding of the data via table summarization, question \& answering, and visual query enabled by \tool.}
    \vspace{-2mm}
    \label{fig:case1}
\end{figure}

As shown in Figure~\ref{fig:case1}, Mike explores the data as follows.

\begin{itemize}
    \item \textbf{Table Summarization.}
    Mike first enters the dialogue box ``\textcolor{grey}{Describe the data.}''.
    The response from \tool includes two parts, one is the basic table description ``\textcolor{teal}{The table shows the market performance of the Dow Jones Index from January 2, 2020, to December 29, 2023, including the date, opening index, intraday high index, intraday low index, and closing index.}'', and the other is the overall trend description ``\textcolor{teal}{The data suffers a sharp drop and then a gradual recovery. There are lots of fluctuations during the overall upward trend.}''.

    \item \textbf{Question \& Answering.}
    Given this description, Mike is interested in the ``sharp drop'' pattern, so he tries to locate the pattern by entering the text ``\textcolor{grey}{Please give me more details about the mentioned `sharp drop' pattern.}''.
    Based on the context, \tool can easily locate the ``sharp drop'' pattern that appears, and the matched interval is highlighted in {\tt Main View}.
   Meanwhile, a text description ``\textcolor{teal}{In March 2020, the Dow Jones Index experienced a `sharp drop' pattern. This drop reflects significant market volatility amid growing concerns over the COVID-19 pandemic.}'' is returned in the dialogue box.

    \item \textbf{Visual Query.}
    Mike then freely sketches a `w' pattern \img{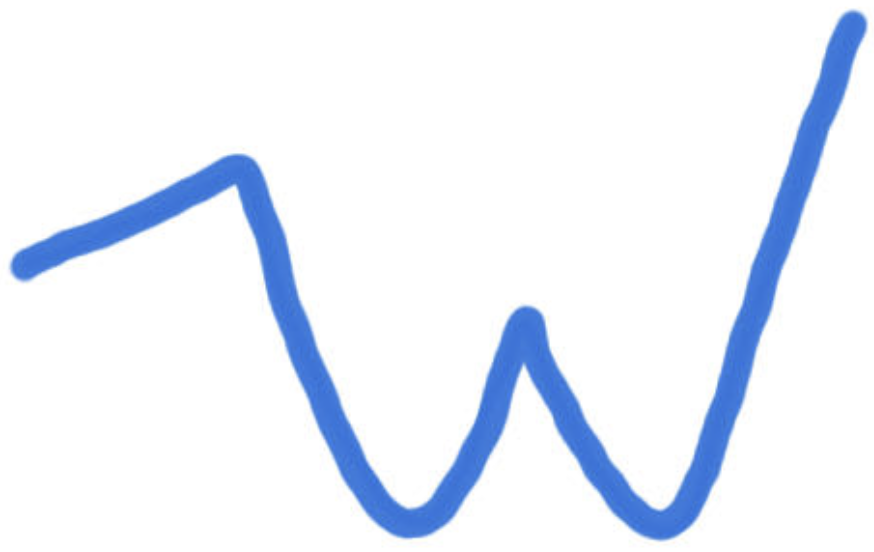} on the empty canvas and enters ``\textcolor{grey}{Are there patterns in the data similar to my hand-drawn sketch?}''.
    \tool \jn{retrieves the} visualization references similar to the input hand-drawn sketch from the vector database, and returns the three most similar matches in {\tt Main View}.
    To facilitate user understanding, \tool also returns a text description of the matched results: ``\textcolor{teal}{The three intervals with similar patterns are from 8-Feb-21 to 16-Mar-21, 18-Feb-21 to 18-Mar-21, and 15-Dec-20 to 11-Jan-21. This pattern is recognized as a double-bottom pattern. It occurred in December 2020 and February 2021, suggesting that during both periods, the market demonstrated buyer interest at lower price levels after experiencing declines. This support prevented further downward movement. Higher prices prompted more buyers to participate, contributing to the Dow's upward momentum.}''.

\end{itemize}

In this way, Mike gets an overall understanding of the tabular data, and also typical trends and patterns within the data.
The multimodal interactions, together with visual interactions like selecting and zooming in the visual interface, enable Mike to easily understand the DJI. data and recognize interesting data change patterns.

\begin{figure*}
    \centering
    \includegraphics[width=0.98\linewidth]{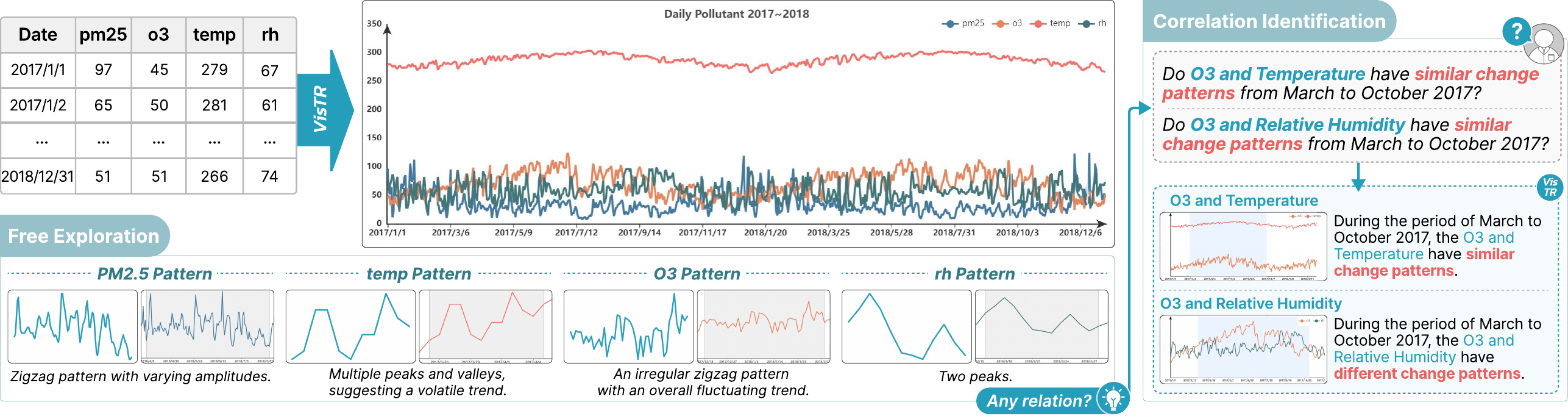}
    \vspace{-2mm}
    \caption{Usage Scenario 2: multivariate analysis for the uploaded air pollutant dataset. With \tool, Emily recognizes typical data change patterns and relationships between different air pollutants.}
    \vspace{-4mm}
    \label{fig:case2}
\end{figure*}

\subsection{Usage Scenario 2: Air Pollutant Analysis}\label{ssec:study2}
In the second usage scenario, we illustrate a hypothetical workflow in which a student, Emily, explores correlations between different air pollutants using \tool.
The used air pollutant dataset \jn{comprises} daily air quality metrics from January 1, 2017, to December 31, 2018, collected by the Beijing U.S. Embassy\footnote{https://aqicn.org/city/beijing/us-embassy/cn/}. 
The data includes variables of Date, PM25, O3, temperature, and relative humidity, which provides insights into the fluctuations in air quality and correlations between air pollutants.
Emily first loads the pollutant data into \tool, which generates a multi-line chart with each line representing a variable.
In the backend, \tool generates a total of 102,787 visualization references, of which 40,243 are stored in Chroma after pruning.

Figure~\ref{fig:case2} illustrates Emily's analysis process using \tool to explore the multivariate data.

\begin{itemize}
    \item \textbf{Free Exploration.}
    To obtain the typical data change patterns of pollutants, Emily selects four pollutant indicators, \emph{PM25, O3, temperature}, and \emph{relative humidity}, and obtains the change patterns in {\tt Pattern View}.
    She notices that all four indicators exhibit obvious daily fluctuations.
    Specifically, PM25 exhibits a `\textcolor{teal}{zigzag pattern with varying amplitudes.}', indicative of its dynamic nature and daily variability.
    Similarly, a typical pattern of O3 is `\textcolor{teal}{An irregular zigzag pattern with an overall fluctuating trend.}', further highlighting the pronounced daily jitters in pollutant levels.
    These observations motivate Emily to explore the relationship between these pollutant indicators over a long time interval.

    \item \textbf{Correlation Identification.}
    Different indicators have different data ranges, making it difficult to directly observe the relationships between different indicators in {\tt Main View}.
    For instance, the value of temperature is always greater than 250 (\textit{K}), while the value of O3 hovers around 50 (\si{\micro\gram\per\cubic\meter}).
    To examine the correlation of these indicators, Emily chooses to directly raise questions to \tool in {\tt Chat Box}.
    She enters ``\textcolor{grey}{Do O3 and temperature have similar change patterns from March to October 2017?}''.
    The O3 and temperature data change patterns during this period are highlighted in {\tt Main View} and the response by \tool is returned in the dialogue box: ``\textcolor{teal}{During the period of March to October 2017, the O3 and Temperature have similar change patterns.}'', and the response indicates a similar pattern between the indicators during the period.
    Emily further asks about the relationship between O3 and relative humidity, by entering ``\textcolor{grey}{Do O3 and Relative Humidity have similar change patterns from March to October 2017?}''.
    From the response, Emily can understand that the data change patterns of O3 and relative humidity are different.
\end{itemize}

In this manner, Emily gains a comprehensive insight into the multivariate air pollutant data, recognizing both the typical data change patterns and correlations among various air pollutant indicators.

\subsection{Evaluation for \tool's Interface}
\label{ssec:interface_feedback}
\noindent
\textit{Participant:} 
We recruited 12 participants (5 females, 7 males) aged 22 to 43, comprising 4 data analysts from industry, 5 Ph.D. students on time-series data analysis, and 3 researchers.
All participants had substantial experience with time-series analysis (Mean=7.25 years) and regularly used data visualization tools in their work.
Before the study, we collected participants' basic information through a questionnaire, including their experience with time-series analysis and visualization tools, familiarity with LLMs, etc.
They \jn{were} invited to participate in the evaluation for the interface and interactions. 
This study qualified for an exemption from ethics board review.
All participants provided informed consent.

\noindent
\textit{Apparatus:} \tool system was deployed on a host equipped with NVIDIA RTX 4090 and a 24-core processor.
This setup \jn{ensured} end-to-end response times under 2 seconds for typical multimodal queries on stored datasets.

\begin{figure}[t]
    \centering
    \includegraphics[width=0.96\linewidth]{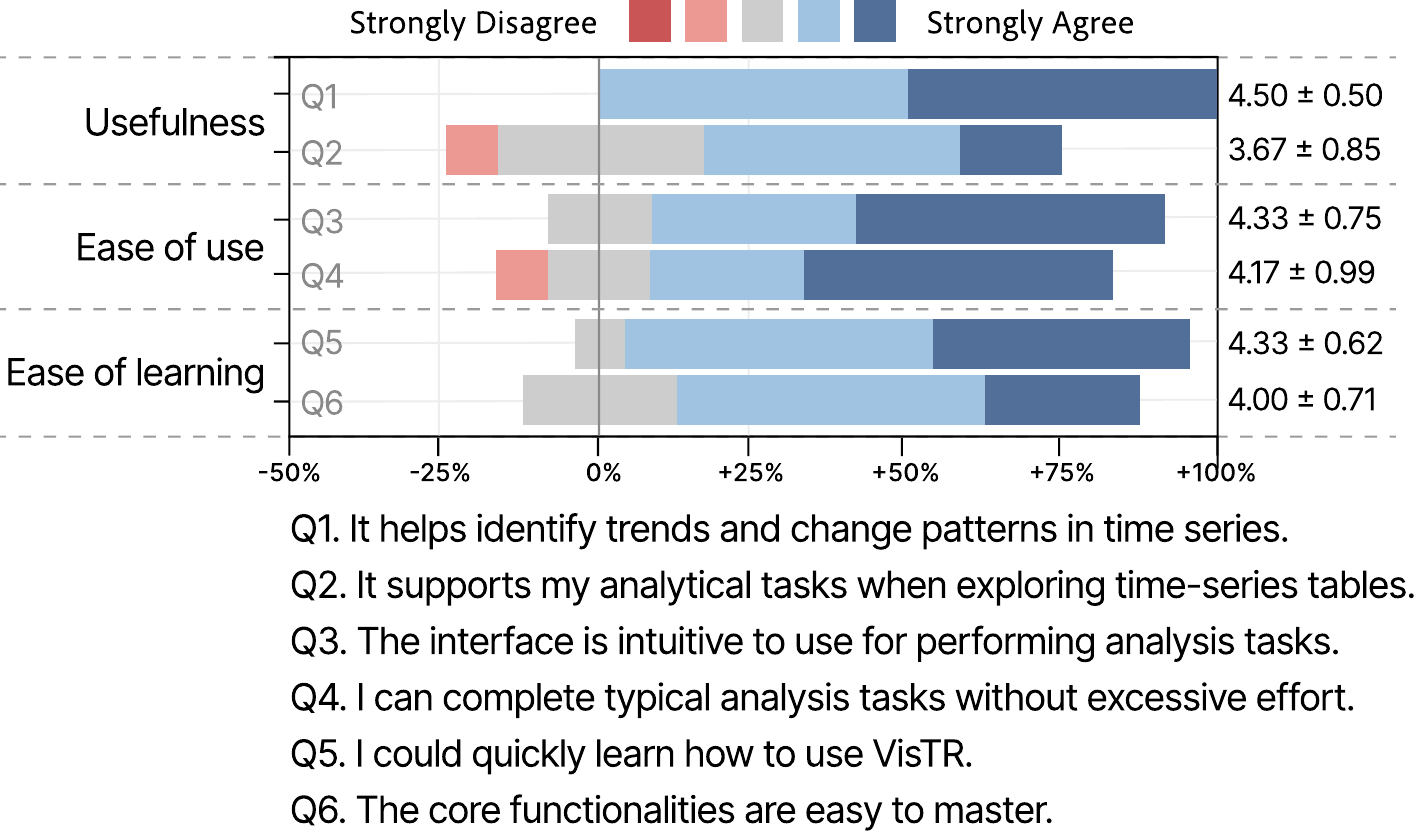}
    \vspace{-2mm}
    \caption{Mean and standard deviation values for usefulness, ease of use, and ease of learning in experiencing \tool.}
    \vspace{-4mm}
    \label{fig:interface_eva}
\end{figure}

\noindent
\textit{Procedure:} 
In this study, we first introduced the proposed framework and its core concepts to participants (20 minutes).
We then walked participants through two representative usage scenarios and instructed participants to familiarize themselves with the interface freely (10-20 minutes). 
They could select a data file and explore it through inputting text, uploading charts, or drawing sketches in {\tt Chat Box}.
For a reference of interest in {\tt Pattern View}, they could click for detailed exploration in {\tt Main View}.
Following the exploration process, we conducted semi-structured interviews (10 minutes) and collected ratings using a post-study questionnaire with a 5-point Likert scale (1=strongly disagree, 5=strongly agree).
The questionnaire assessed three aspects: usefulness, ease of use, and ease of learning.
Each aspect was measured using two Likert items.
Specifically, usefulness items focused on whether the interface effectively \jn{supported} time-series exploration tasks, especially for identifying long-term trends and patterns (Q1-Q2 in Figure~\ref{fig:interface_eva}).
Ease of use items assessed the perceived intuitiveness and effort required to perform analytical tasks (Q3-Q4).
Ease of learning items evaluated the interface’s accessibility, focusing on how quickly participants believed they could become proficient with the \tool interface and its core functionalities (Q5-Q6).
Participants were also encouraged to discuss any difficulties they encountered and to suggest potential improvements to \tool.
Each study session lasted approximately 40-50 minutes per participant.

\noindent
\textit{Results:}
Participant ratings are shown in Figure~\ref{fig:interface_eva}.
Overall, \tool received positive ratings for usefulness (Mean=4.08, SD=0.49), ease of use (Mean=4.17, SD=0.69), and ease of learning (Mean=4.17, SD=0.55).
Notably, for the system's usefulness, each participant rated Q1 above 4, indicating strong agreement that \tool effectively \jn{supported} time-series change pattern exploration tasks.
In addition to the questionnaire, we analyzed feedback from the semi-structured interviews.
Overall, participants reported that multi-modal interaction enabled more intuitive and interesting expression of exploring time-series change patterns.
Several participants noted that combining natural language input with sketching helped them refine vague ideas into more precise queries, while {\tt Pattern View} was appreciated for providing an overview of typical change patterns and supporting exploratory reasoning.
For example, one participant with extensive expertise in time-series analysis mentioned that her traditional analytical workflow typically \jn{involved} using Excel to plot overall trends or Tableau to generate dashboards before conducting manual data mining.
She found that \tool could streamline this process, enabling rapid insight discovery from high-level overviews to in-depth insights.
Another participant, a specialist in financial time series analysis, highlighted \tool's robust performance on line charts, for displaying stock moving averages.
As a next step, he suggested extending \tool to support domain-specific visual representations like candlestick charts.
Integrating these charts with supplementary market reports could realize automated market analysis and pattern reasoning.
Additionally, some participants suggested potential improvements, such as integrating SQL query capabilities into the framework to better support large-scale table exploration without requiring programming expertise.

\section{Discussion}\label{sec:discussion}

By leveraging visualizations as representations, \tool successfully addresses the limitations of existing LLM-based table reasoning methods in pattern recognition and visual-based pattern exploration.
Moreover, the process also aligns with exploratory data analysis (EDA) process, which often involves summarizing the main characteristics of a dataset using visualization, enabling users to understand the data structure, recognize patterns, detect anomalies, and formulate hypotheses for further investigation.
In this sense, the proposed framework not only streamlines the EDA process but also expands its horizons, providing users with novel approaches to uncover hidden patterns within datasets.

\stitle{Generalizability.} Nevertheless, to fully realize the potential of the proposed framework, we need to enhance the generalizability of \tool from the following perspectives.

\begin{itemize}[leftmargin=*]
\item \emph{More table types.} 
    \tool is currently instantiated on time-series tables, which contain data change patterns that \jn{may not be readily} processed by existing SQL-based LLMs for table reasoning.
    \tool has the potential to be extended to more general table types by employing alternative data transformation methods within the \emph{visualization referencing} module.
    For instance, \tool can be adapted to recognize and analyze the distribution patterns in a student grade table.
    Supporting a broader range of table types is feasible, leveraging methods such as subset embedding\cite{sen_2021} and learning-based approaches\cite{sun2022learning}. 

    \item \emph{General queries.}
    \tool focuses on tasks related to reasoning about data change patterns in tables.
    Therefore, we primarily provide text-based user queries related to data change patterns, such as \q{what type of change pattern exists in the data?}
    However, this narrow focus also limits its ability to handle general user queries, including those related to causal reasoning, such as \q{why did Apple stock rise in March?}
    When posed with such questions, \tool may give incorrect queries and responses.
    Moreover, due to the stochastic nature of the base LLM, \tool may generate different outputs across multiple runs, even with the same input.
    This inherent variability presents challenges for applications requiring consistent answers. 
    To \jn{enable} comprehensive exploration and reasoning \jn{over} a data table, more general user queries \jn{are required}, which \jn{could benefit from} the incorporation of external knowledge sources to ensure consistency and accuracy.

    \item \emph{Dataset bias.}
    The successful alignment of our model heavily relies on acquiring high-quality training data.
    However, the dataset used for fine-tuning, while demonstrating its effectiveness, is limited in size and diversity. 
    Scaling up the dataset to address these limitations presents significant challenges.
    First, the limited number of users contributing to user-labeled sketches may embed subjective preferences into the chart-sketch-text pairings.
    While crowd-sourcing strategies offer a way to accelerate data collection, they \jn{can be susceptible to} subjective biases, and quality control for such large-scale manual efforts \jn{may remain} a significant challenge.
    An alternative lies in data augmentation techniques to reduce reliance on manual sketch collection.
    Second, the filtered trend words from textual descriptions may oversimplify visual information, neglecting nuances in semantics.
    To enrich textual descriptions, a potential solution proposed by Time-LLM\cite{timellm_2024} is to generate text prototypes by assembling candidate vocabularies.  
    Then the generated rich text descriptions can be paired with charts through crowd-sourced rating frameworks\cite{difference_2023}.

    \item \emph{Intent-enhanced sketches.}
    Users' intuition about sketching targets often emerges from the exploration process.
    For this reason, \tool offers insights into common patterns within the uploaded table.
    However, expressing complex or uncommon patterns solely through hand-drawn sketches can be challenging for users\cite{Lee_2020_visualquery}. 
    To address this challenge, several solutions can be considered.
    First, incorporating additional interactive features like dragging, resizing, and erasing can provide users with greater flexibility in adjusting their intent~\cite{inksight_2023}.
    Summarizing multi-attribute temporal patterns and introducing a more effective time-series segmentation method have shown promise\cite{SHIRATO202377}. 
    By implementing these methods, we can further enhance the capabilities of \tool to support sketches with enhanced intents.

\end{itemize}

\stitle{VisTR vs. Traditional Methods.}
\tool introduces an innovative attempt to convert time-series tables into visual representations, enabling efficient multi-modal alignment and continuous context reasoning.
Compared to traditional subsequence or sketch matching methods, such as metric-based\cite{qetch_2018}, regular expression-based\cite{saxregex_2023} or deep-learning-based approaches\cite{LSTM_sketch_2020}, \tool processes varying data facet lengths into uniform vision tokens, offering smoother multi-modal interaction and resource savings \jn{from} reduced prompt lengths\cite{plotsts_2024}. 
However, visual representations come with notable limitations. 
They often underperform traditional metric-based methods in matching precision, incur higher storage costs, and may ignore shorter time subsequences in the \emph{visualization referencing} module.
Additionally, fine-tuning models for aligning modalities of chart, sketch, and text is resource-intensive.
Despite these challenges, the rapid advancements of MLLMs make this research direction promising.
Future efforts should focus on improving precision and optimizing storage efficiency to fully leverage the potential of visual representations in advancing multimodal frameworks.

\stitle{Potential scenarios.}
Besides table reasoning, the proposed framework holds potential for other application scenarios, such as safeguarding sensitive data that requires a balance between data privacy and utility.
By transforming raw tabular data into visualizations, \tool shields the secure information contained within tables from users, while still allowing for examination and querying of the data patterns.
Moreover, \tool could also find applications in fraud detection, as visualizations generated from transactional data could help recognize suspicious patterns indicative of fraudulent activities.

\stitle{Lack of Evaluation Datasets.}
A notable limitation in time-series LLMs is the \jn{scarcity} of paired time series-text-chart data and robust evaluation methodologies~\cite{chatts_2024}.
Multimodal datasets that align long-term time-series data with corresponding textual descriptions and charts would facilitate the evaluation of time-series understanding and reasoning tasks, \jn{while also} enabling advanced tasks such as insight mining and chart captioning~\cite{Vistext_2023}.
\jn{The construction of} such datasets \jn{may help} bridge the gap between raw time-series processing and human-centric insights, \jn{contributing to} more effective multimodal time-series intelligence.

\section{Conclusion}
We present \tool, a new framework for time-series table reasoning that leverages visualizations as \emph{representations} to bridge raw data and human cognition. By enabling enhanced pattern recognition, mitigating \jn{context loss}, and supporting visual-based exploration, \tool addresses key limitations of existing LLM-based methods.
\tool integrates a fine-tuned MLLM, trained on a newly constructed dataset of chart-text and chart-sketch pairs, to align charts, text, and sketches in a unified embedding space. Combined with strategies for visualization referencing, pruning, and indexing, \tool provides an interactive system for seamless multimodal interactions with time-series tables.
Quantitative evaluations demonstrate its effectiveness in aligning multimodal inputs and improving reasoning accuracy, while usage scenarios and user feedback together prove its usability.
\tool opens new research directions for integrating visualizations into table reasoning and enhancing multimodal data exploration. Future work will extend its applicability to broader table types, generalized queries, and diverse application scenarios.

\section*{Acknowledgment}
The authors would like to thank the editors and reviewers for their valuable comments and suggestions.
This work was supported in part by the National Natural Science Foundation of China (62572415, 62372321, 62402409), 
Youth S\&T Talent Support Programme of Guangdong Provincial Association for Science and Technology (SKXRC2025461), 
and the Tianjin Natural Science Foundation (25JCZDJC01270).

\bibliographystyle{IEEEtran}
\bibliography{references}

\end{document}